# High-$T_c$ superconductivity originated from strong spin-charge correlation: indication from linear temperature dependence of resistivity

Tian-De Cao

*Department of physics, Nanjing University of Information Science & Technology, Nanjing 210044, China*

**Both the highest-$T_c$ and the linear temperature dependence of the resistivity in wide temperature range appear at the optimally doped regions of Cu-based superconductors[1,2,3,4,5], and the highest-$T_c$ of Fe-based superconductors[6,7] are also associated with the linear temperature dependence of the resistivity in normal states near superconducting states. This means that the high temperature superconductivity and the linear temperature dependence of the resistivity should be dominated by the same mechanism. This letter on theoretic calculation clearly shows that strong spin-charge correlation dominated resistivity behaves the linear temperature dependence, thus high-temperature superconductivity should be induced by strong spin-charge correlation.**

Let us consider the extended Hubbard model

$$H = \sum_{l,l',\sigma}(t_{ll'}-\mu\delta_{ll'})d^+_{l\sigma}d_{l'\sigma} + U\sum_l n_{l\sigma}n_{l\bar\sigma}$$
$$+\frac{1}{4}\sum_{l,l',\sigma,\sigma'}V_{ll'}n_{l\sigma}n_{l'\sigma'} - \sum_{l,l'}J_{ll'}\hat{S}_{lz}\hat{S}_{l'z} \qquad (1)$$

Where we considered the charge-charge interaction and the spin-spin interaction, while the terms including $\hat{S}_x(\hat{S}_y)$ are neglected. After introducing the charge operator $\hat\rho(q)=\frac{1}{2}\sum_{k,\sigma}d^+_{k+q\sigma}d_{k\sigma}$ and the spin operator $\hat{S}(q)=\frac{1}{2}\sum_{k,\sigma}\sigma d^+_{k+q\sigma}d_{k\sigma}$ in the wave vector space, we write the model (1) in

$$H_0 = \sum_{k,\sigma}\xi_k c^+_{k\sigma}c_{k\sigma} + \sum_q V(q)\hat\rho(q)\hat\rho(-q)$$
$$-\sum_q J(q)\hat{S}_z(q)\hat{S}_z(-q) \qquad (2)$$

Define Green's function

$$G(k\sigma,\tau-\tau') = -<T_\tau c_{k\sigma}(\tau)c^+_{k\sigma}(\tau')> \qquad (3)$$

and establish the equation of this function, we shall calculate $\partial_\tau G$; to consider higher level of approximation, we have to calculate $\partial_\tau <T_\tau \hat\rho(q)c_{k+q\sigma}c^+_{k\sigma}(\tau')>$, $\partial_\tau <T_\tau \hat{S}(q)c_{k+q\sigma}c^+_{k\sigma}(\tau')>$, and so on. The final equation is arrived at

$$[-i\omega_n + \tilde\xi_{k\sigma} + \sum_q \frac{P(k,q,\sigma)}{i\omega_n - \xi_{k+q}}]G(k\sigma,i\omega_n)$$
$$= -1 + \frac{I_\sigma}{-i\omega_n + \xi_k} \qquad (4)$$

Where

$$\tilde\xi_{k\sigma} = \xi_k - \sigma\bar{s}_{lz}J(0) - \sum_q [V(q)-J(q)]n_{k+q,\sigma}/2$$

$$I_\sigma = \sigma J(0)\bar{s}_{lz} + V(0)n_l/2 \qquad (5)$$

$$P(k,q,\sigma) = \frac{1}{2}(\xi_{k+q}-\xi_k)(-J(q)+V(q))n_{k+q,\sigma}$$
$$+ J(-q)P_{ss}(q,\tau=0)J(q)$$
$$-2\sigma V(-q)P_{s\rho}(q,\tau=0)J(q)$$

$$+V(-q)P_{\rho\rho}(q,\tau=0)V(q)$$

Where $P_{\rho\rho}$, $P_{ss}$ and $P_{s\rho}$ are correlation functions at the same time, $\bar{s}_{lz}$ is the spin at each site. We rewrite Eq. (4) in

$$G(k\sigma,i\omega) = \frac{1}{i\omega_n - \tilde{\xi}_{k\sigma} - \Sigma(k\sigma,i\omega_n)} \quad (6)$$

and express the self-energy as

$$\Sigma_{ret} = \mathrm{Re}\,\Sigma_{ret}(k\sigma,\omega) + i\,\mathrm{Im}\,\Sigma_{ret}(k\sigma,\omega) \quad (7)$$

$$\mathrm{Im}\,\Sigma_{ret}(k\sigma,\omega) =$$
$$-\pi[-I_\sigma \sum_q P(k,q,\sigma)/(\omega - \xi_{k+q}) + \omega - \tilde{\xi}_{k\sigma}]\delta(\omega - \xi_k + I_\sigma)$$
$$-\pi(\omega - \xi_k)\sum_q P(k,q,\sigma)\delta(\omega - \xi_{k+q})/(\omega - \xi_k + I_\sigma)$$
$$(8)$$

The form of $\mathrm{Re}\,\Sigma_{ret}$ can be given if it were necessary. The temperature dependence of resistivity can be found in $\mathrm{Im}\,\Sigma_{ret}$, but the temperature dependence of $\mathrm{Im}\,\Sigma_{ret}$ is dominated by $P(k,q,\sigma)$, and the temperature dependence of $P(k,q,\sigma)$ is dominated by these correlation functions $P_{\rho\rho}$, $P_{ss}$ and $P_{s\rho}$, which can be seen in Eqs. (5).

To calculate $\mathrm{Im}\,\Sigma_{ret}$, we must find functions $P_{\rho\rho}$, $P_{ss}$ and $P_{s\rho}$. These correlation functions are defined as

$$P_{\rho\rho}(q,\tau-\tau') = <T_\tau \hat{\rho}(q,\tau)\hat{\rho}(-q,\tau')>$$
$$P_{s\rho}(q,\tau-\tau') = <T_\tau \hat{S}(q,\tau)\hat{\rho}(-q,\tau')> \quad (9)$$
$$P_{ss}(q,\tau-\tau') = <T_\tau \hat{S}(q,\tau)\hat{S}(-q,\tau')>$$

To find these correlation functions, we first calculate $\partial_\tau P_{\rho\rho}$, $\partial_\tau P_{ss}$ and $\partial_\tau P_{s\rho}$, then calculate $\partial_\tau <T_\tau c^+_{k+q\sigma}c_{k\sigma}c^+_{k'-q\sigma'}(\tau')c_{k'\sigma'}(\tau')>$ to higher level of approximation. After lengthy calculating, we obtain

$$P_{\rho\rho}(q,\tau=0) = -\sum_{k,k',\sigma,\sigma'} \frac{1}{32\pi^2}(\xi_{k+q} - \xi_k)$$
$$\cdot \int_{-\infty}^{+\infty} d\varepsilon d\eta [n_B(\eta)n_F(\varepsilon) + n_B(\varepsilon)n_F(\eta)]\frac{1}{\varepsilon - \eta}\frac{[\sigma\sigma' J(q) - V(q)]}{(\varepsilon - \eta + \xi_{k+q} - \xi_k)}$$
$$\cdot [n_{k\sigma} - n_{k+q\sigma}] A(k'\sigma',\eta)A(k'-q\sigma',\varepsilon) \quad (10)$$

The forms of both $P_{ss}$ and $P_{s\rho}$ are similar to this expression. The spectral function in these correlation functions are dominated by the energy region of $\mathrm{Im}\,\Sigma_{ret} = 0$, thus we take this form

$$A(k'\sigma',\eta) = z_{k'\sigma'}\delta(\eta - E_{k'\sigma'}),$$ and obtain

$$P_{ss}(q \neq 0, \tau = 0)$$
$$= J(q)\sum_{\sigma\sigma'} p_{\sigma\sigma'}(q,T) - V(q)\sum_{\sigma\sigma'}\sigma\sigma' p_{\sigma\sigma'}(q,T)$$

$$P_{\rho\rho}(q \neq 0, \tau = 0)$$
$$= J(q)\sum_{\sigma\sigma'}\sigma\sigma' p_{\sigma\sigma'}(q,T) - V(q)\sum_{\sigma\sigma'} p_{\sigma\sigma'}(q,T) \quad (11)$$

$$P_{s\rho}(q \neq 0, \tau = 0)$$
$$= J(q)\sum_{\sigma\sigma'}\sigma' p_{\sigma\sigma'}(q,T) - V(q)\sum_{\sigma\sigma'}\sigma p_{\sigma\sigma'}(q,T)$$

where we introduced

$$p_{\sigma\sigma'}(q,T) = -\sum_{k,k'} \frac{1}{32\pi^2}(\xi_{k+q} - \xi_k)$$
$$\cdot \frac{1}{E_{k'-q,\sigma'} - E_{k'\sigma'}}\frac{1}{(E_{k'-q\sigma'} - E_{k'\sigma'} + \xi_{k+q} - \xi_k)}$$
$$\cdot [n_{k\sigma} - n_{k+q\sigma}] z_{k'\sigma'} z_{k'-q\sigma'}$$
$$\cdot [n_B(E_{k'\sigma'})n_F(E_{k'-q,\sigma'}) + n_B(E_{k'-q,\sigma'})n_F(E_{k'\sigma'})] \quad (12)$$

Include the terms of $q = 0$, we have

$$P_{ss}(q, \tau = 0) = \bar{s}_{lz}\bar{s}_{lz}\delta_{q,0} + P_{ss}(q \neq 0, \tau = 0)$$

$$P_{\rho\rho}(q,\tau=0) = \frac{1}{4}n_l n_l \delta_{q,0} + P_{\rho\rho}(q \neq 0, \tau=0)$$

$$P_{s\rho}(q,\tau=0) = \frac{1}{2}\bar{s}_{lz} n_l \delta_{q,0} + P_{s\rho}(q \neq 0, \tau=0)$$

For non-ferromagnetic states, the spin $\bar{s}_{lz}=0$ at each site. These correlation functions and the energies $E_{k\sigma}$ should be determined consistently, and this seems very complex. However, our focus is on the temperature dependence of resistivity, and we find the temperature dependence of $p_{\sigma\sigma'}(q,T)$ is $p_{\sigma\sigma'} \propto T$ because $p_{\sigma\sigma'}$ is dominated by $n_B(E_{k\sigma})$ around $E_{k\sigma} \sim 0$, $n_B(E_{k\sigma}) \sim T/E_{k\sigma}$, and this leads $\text{Im}\Sigma_{ret} \propto T$. That is, the correlations dominated resistivity is $\rho \propto T$. However, the $\rho \propto T$ behavior requires the "strong correlation", where the temperature dependence of $n_{k+q,\sigma}$ can be neglected in Eqs.(5). What is the "strong correlation"? It is seen with Eq. (11) that the spin-spin correlation is dominated by the exchange coupling $J(q)$, the charge-charge correlation is dominated by $V(q)$, and the spin-charge correlation is dominated by both $J(q)$ and $V(q)$ ( the spin-charge correlation has been neglected by other authors in all literatures). The effects of the spin-spin correlation is "enlarged" by $J(q)$, the effects of the charge-charge correlation is "enlarged" by $V(q)$, and the effects of the spin-charge correlation is "enlarged" by $V(q)$ and $J(q)$, which can be found in Eqs. (5). Since $V(q) = U + V_0(q)$ and $J(q) = U + J_0(q)$, it is seen that the strong correlation could be symbolized by $U$ the on-site interaction (which has been suggested by other physicists, but this work gives an evident expression). It is necessary to note that the strong correlation requires that both the spin correlation and the charge correlation are strong, and this is consistent with our previous suggestion[8]. This is to say, the strong correlation can be stood for by the spin-charge correlation. Because the strong correlation usually corresponds to both spin excitations and charge excitations, thus the superconducting pairing should be mediated by spin excitations and charge excitations[9]. We must emphasize that effects of other factors (phonons, impurities, hopping between bands, and so on) on resistivity should be considered for weak correlation systems, which will lead to other temperature dependences of resistivity.

Particularly, the strong correlation dominated resistivity can behave $\rho \propto T$ in a wide temperature range as shown in the calculation above, if only the strong correlation can be kept.

In the Hamiltonian (1) above, we do not consider the electron-phonon interaction, the electron-impurity interaction, and so on. It is well-known that the high temperature superconductivity usually occurs in the strong correlation materials, and the temperature dependence of resistivity in normal state of these materials is usually linear, thus that the high temperature superconductivity is induced by the strong correlation seems an inevitable conclusion.